\documentclass{elsart}                                               
\usepackage{amssymb}
\journal{Physics Letters B}
\begin{document}
\begin{frontmatter}

\title{Integrable Structure of Superconformal
Field Theory and Quantum super-KdV Theory}
\author[POMI]{Petr P. Kulish},
\ead{kulish@pdmi.ras.ru}
\author[SpbU]{Anton M. Zeitlin}
\ead{zam@math.ipme.ru}
\address[POMI]{St. Petersburg Department of Steklov Mathematical 
Institute, Fontanka, 27,\\ St. Petersburg, 191023, Russia}
\address[SpbU]
  {Department of High Energy Physics, Physics Faculty, St. Petersburg State 
  University, Ul'yanovskaja 1, Petrodvoretz, St.Petersburg, 198904, Russia}

\begin{abstract}
The integrable structure of the two dimensional superconformal field theory  
is considered. The classical
counterpart of our constructions is based on the $\widehat{osp}(1|2)$ 
super-KdV hierarchy. 
The quantum version of the monodromy matrix associated with the 
linear problem for the corresponding L-operator is introduced.
Using the explicit form of the irreducible representations of
$\widehat{osp}_q(1|2)$, 
the so-called ``fusion relations'' for the transfer
matrices considered in different representations of 
$\widehat{osp}_q(1|2)$ are obtained.
The possible integrable perturbations 
of the model (primary operators, commuting
with integrals of motion) are classified
and the relation with the supersymmetric $\widehat{osp}(1|2)$ Toda
field theory is discussed.
\end{abstract}
\end{frontmatter}
\section{Introduction}
Conformal field theory (CFT) provides effective 
tools to classify the fields in 
the theory we study and to compute their correlation functions. 
Perturbartion leads the system out of the critical point and breaks
the conformal 
invariance. But special perturbations, called
''integrable'' still preserve an infinite-dimensional abelian 
algebra of conserved 
charges, thus 
leading to an integrable theory.\\
\hspace*{5mm} The authors of \cite{1}, \cite{2} showed that in this case 
the problem could be studied 
from a point of view of continuous field theory version of the quantum inverse 
scattering method (QISM) \cite{leshouches},\cite{sklyan}. 
The proposition is the following: 
at first, to use CFT symmetries for construction of QISM structures at 
the scale invariant fixed point and then to study the integrable perturbed 
model by obtained QISM tools.
 Our object of study is a model based on the 
superconformal symmetry (see e.g. \cite{SCFT1},\cite{SCFT2}).
The superconformal field theory has been applied 
to the physics of 2D disordered systems, 
to the study of lattice models (the tricritical 
Ising model) and to the superstring physics. 
We build basic 
structures of QISM: quantum monodromy matrices, RTT-relation and 
fusion relations for the transfer matrices,
using underlying superconformal symmetry. \\
\hspace*{5mm}The  $\widehat{osp}(1|2)$  
supersymmetric Korteweg - de Vries theory (super-KdV) 
\cite {sKdV1}-\cite {sKdV4} is used as a classical limit of our
quantum system. Due to the Drinfeld-Sokolov
reduction of the $\widehat {osp}(1|2)$
affine superalgebra, Miura transformation and 
Poisson brackets are introduced. 
Then the monodromy matrix is constructed. The associated
auxiliary L-matrix satisfies the r-matrix quadratic Poisson
bracket relation and plays a crucial role in the following.\\
\hspace*{5mm}After this necessary preparation,  
we move to the quantum theory (Sec. 3). The quantum 
Miura transformation is realized by 
the free field representation of the 
superconformal algebra \cite{SCFT1},\cite{SCFT2}. Then the 
super-Virasoro module is considered,
where the quantum versions of integrals of motion (IM) act.
Vertex operators  are also introduced,
to build quantum monodromy matrix.\\
\hspace*{5mm}The algebraic structure corresponding to the quantum case, 
coincides 
with the quantum superalgebra $\widehat {osp}_q(1|2)$.
We construct the finite dimensional irreducible representations of 
this quantum Lie superalgebra (Sec. 4).
It appears, that in 
quantum monodromy matrix (in comparison with 
the classical case) one term 
is missing in the P-exponent.
The quantum L-matrix satisfies the so-called RTT-relation, giving the
integrability condition in the quantum case. 
Considering monodromy
matrices in the obtained $\widehat {osp}_q(1|2)$ representations,
we get the functional relations (``fusion relations'') for their traces --
``transfer matrices''.
When the deformation parameter is rational (the case of CFT minimal
models), these fusion relations become the closed system of equations,
 which, due to the conjecture of \cite{1} can be used to find the
full set of eigenvalues of transfer matrices. Also, we
suppose that they could be transformed to the Thermodynamic Bethe
Ansatz equations \cite{TBA}.\\
\hspace*{5mm}Finally, we
discuss integrable perturbations of the model and the relation with
the supersymmetric Toda field theory. 
\section{A review of classical super-KdV theory} 
The classical limit of constructions of papers \cite{1}, \cite{2} 
leads to the 
Drinfeld-Sokolov KdV hierarchies related to the corresponding affine 
Lie algebras 
$A_1^{(1)}$ and $A_2^{(2)}$. Our quantum model gives in classical limit
the super-KdV hierarchy \cite{sKdV1}-\cite{sKdV4} related 
to $B(0,1)^{(1)}$ (or 
$\widehat {osp}(1|2)$) affine Lie superalgebra. 
 The supermatrix L-operator, corresponding to super-KdV theory
is the following one:
\begin{eqnarray}
\mathcal{L}_F=D_{u,\theta} 
-D_{u,\theta}\Psi h-(iv_{+} \sqrt{\lambda}-\theta\lambda X_{-}),
\end{eqnarray}
where $D_{u,\theta} =\partial_\theta + \theta \partial_u$ 
is a superderivative, variable 
$u$ lies on a cylinder of circumference $2\pi$, $\theta$ 
is  a  Grassmann  variable, $\Psi(u,\theta)=\phi(u) - 
i \theta\xi(u)/\sqrt{2}$ 
is a bosonic superfield; $h,v_+,v_-,X_-,X_+$ are generators of 
$osp(1|2)$ (for more information see \cite{osp1}, \cite{osp2}):
\begin{eqnarray}
[h,X_\pm]&=&\pm 2X_\pm ,\quad [h,v_\pm]=\pm v_\pm ,\quad [X_+,X_-]=h,\\
\lbrack v_\pm,v_\pm\rbrack&=&\pm 2 X_\pm, \quad  [v_{+},v_{-}]=-h, \quad
[X_\pm,v_\mp]= v_\pm, \quad [X_\pm,v_\pm]=0.\nonumber
\end{eqnarray}
Here [,] means supercommutator: $[a,b]=ab-(-1)^{p(a)p(b)}ba$
and the parity $p$ is defined as follows:
$p (v_\pm)=1$, $p (X_\pm)=0$, $p (h)=0$. The 
``fermionic'' operator $\mathcal{L}_F$ considered together 
with a linear problem      
$\mathcal{L}_F\chi(u,\theta)=0$ is equivalent to the ``bosonic'' one:
\begin{eqnarray}
\mathcal{L}_B=\partial_u-\phi'(u)h-\sqrt{\lambda/2}\xi(u)v_{+}-\lambda (X_{+}+X_{-}).
\end{eqnarray}
The fields $\phi$, $\xi$ satisfy the following boundary conditions:
\begin{eqnarray}
\phi(u + 2\pi )=\phi(u)+ 2\pi i p,\quad
\xi(u + 2\pi )=\pm\xi(u), 
\end{eqnarray}
where ``+'' corresponds to the so-called Ramond (R) sector of the model and   
``--'' to the Neveu-Schwarz (NS) one.
 The Poisson brackets, given by the Drinfeld-Sokolov construction are the
following:
\begin{eqnarray}
\{\xi(u),\xi(v)\}=-2\delta(u-v),\quad
\{\phi(u),\phi(v)\}=\frac{1}{2}\epsilon(u-v).
\end{eqnarray}
The L-operators (1), (3) correspond to the super-modified KdV, they 
are written in the 
Miura form. Making a gauge
transformation to proceed to the super-KdV L-operator one obtains two fields:
 \begin{eqnarray}\label{eq:Miura}
U(u)=-\phi''(u)-\phi'^2(u)-\frac{1}{2}\xi(u)\xi'(u),\quad
\alpha(u)=\xi'(u) + \xi(u)\phi'(u),
\end{eqnarray}
which generate the superconformal algebra under the 
Poisson brackets:
\begin{eqnarray}
\{U(u),U(v)\}&=&
 \delta'''(u-v)+2U'(u)\delta(u-v)+4U(u)\delta'(u-v),\\
\{U(u),\alpha(v)\}&=&
 3\alpha(u)\delta'(u-v) + \alpha'(u)\delta(u-v),\nonumber\\
\{\alpha(u),\alpha(v)\}&=&
 2\delta''(u-v)+2U(u)\delta(u-v).\nonumber
\end{eqnarray} 
These brackets describe the second Hamiltonian structure of the super-KdV 
hierarchy. One can obtain an evolution equation by taking one of the 
corresponding infinite set of local IM
(they could be obtained by expanding log$(\mathbf{t}_1(\lambda))$, where
$\mathbf{t}_1(\lambda)$ is the supertrace of the monodromy matrix, see below):
\begin{eqnarray}
&&I^{(cl)}_1=\int U(u)\d u,\quad
I^{(cl)}_3=\int\Big(\frac{U^2(u)}{2}+\alpha(u)\alpha'(u)\Big)\d u,\\
&&I^{(cl)}_5=\int\Big((U')^2(u)-2U^3(u)+8\alpha'(u)\alpha''(u)+12
\alpha'(u)\alpha(u)U(u)\Big)\d u,\nonumber\\
&&  .\qquad.\qquad.\nonumber
\end{eqnarray}
These conservation laws form an involutive set under the Poisson brackets:
$
\{I^{(cl)}_{2k-1},I^{(cl)}_{2l-1}\}=0.
$
From the $I^{(cl)}_3$ conservation law one obtains the super-KdV equation
for the Grassmann algebra valued functions
\cite{sKdV1}-\cite{sKdV3}:
\begin{eqnarray}
U_t=-U_{uuu}-6UU_u - 6\alpha\alpha_{uu},\quad
\alpha_t=-4\alpha_{uuu}-6U\alpha_u -3U_u\alpha.
\end{eqnarray}
Now let's consider the ``bosonic'' linear problem 
$\pi_s(\mathcal L_B)\chi(u)=0$,
where $\pi_s$ means irreducible representation of $osp(1|2)$ 
labeled by an integer $s\ge 0$ \cite{osp1},\cite{osp2}. 
We can write the solution of this problem in such a way:
\begin{eqnarray}
\chi(u)&=&\pi_s(\lambda)
\Bigg(e^{-\phi(u)h_{\alpha_0}}P\exp\int_0^u \d u'\Big(\xi(u')e^{-\phi(u')}e_{\alpha}
+e^{-2\phi(u')}2e_{\alpha}^2 \\
&+& e^{2\phi(u')}e_{\alpha_0}\Big)\Bigg)\chi_0,\nonumber
\end{eqnarray}
where $P\exp$ means $P$-ordered exponent, $\chi_0 \in C^{2s+1}$ 
is a constant 
vector and $e_{\alpha}$, $ e_{\alpha_0}$, $h_{\alpha_0}$ 
are the Chevalley generators of $\widehat {osp}(1|2)$ (see \cite{Tolst}),
which coincide in the evaluation representations $\pi_s(\lambda)$
with $\sqrt{\lambda/2}v_+$, ${\lambda}X_-$, $-h$ correspondingly.  
The associated monodromy matrix then has the form:
\begin{eqnarray}
\mathbf{M}_s(\lambda)&=&\pi_s\Bigg(e^{-2\pi iph_{\alpha_0}}
P\exp\int_0^{2\pi} \d u\Big(\xi(u)e^{-\phi(u)}e_{\alpha}
+e^{-2\phi(u)}2e_{\alpha}^2 \\
&+& e^{2\phi(u)}e_{\alpha_0}\Big)\Bigg).\nonumber
\end{eqnarray}
Following \cite{1} let's introduce auxiliary matrices:
$
\pi_s(\lambda)(\mathbf{L})=\mathbf{L}_s(\lambda)=\\
\pi_s(\lambda)(e^{\pi ip h_{\alpha_0}})
\mathbf{M}_s(\lambda).
$
They satisfy Poisson bracket algebra \cite{Fadd}:
\begin{eqnarray}
\{\mathbf{L}_s(\lambda)\otimes_{,}\mathbf{L}_{s'}(\mu)\}=[\mathbf{r}_{ss'}(\lambda\mu^{-1}),\mathbf{L}_{s}(\lambda)\otimes \mathbf{L}_{s'}(\mu)],
\end{eqnarray}
where $\mathbf{r}_{ss'}(\lambda\mu^{-1})=\pi_s(\lambda)
\otimes\pi_{s'}(\mu)(\mathbf{r})$ 
is the classical trigonometric $\widehat{osp}(1|2)$
r-matrix \cite{Kulish} taken in the corresponding representations.
From the Poisson brackets for $\mathbf{L}_s(\lambda)$ one obtains that the 
traces
of monodromy matrices $\mathbf{t}_s(\lambda)=str\mathbf{M}_s(\lambda)$ 
commute under the Poisson
bracket: $
\{\mathbf{t}_s(\lambda),\mathbf{t}_{s'}(\mu)\}=0$. If one expands 
log($\mathbf{t}_1(\lambda)$) in the $\lambda^{-1}$ power series, one can see 
that the coefficients in this expansion are the local IM, as we mentioned earlier.
\section{Free field representation of Superconformal algebra
and Vertex operators}
To quantize the introduced 
classical quantities, we start from a quantum version 
of the Miura transformation (\ref{eq:Miura}),
the so-called free field 
representation of the superconformal algebra \cite{SCFT1}:
\begin{eqnarray}
-\beta^2T(u)&=&:\phi'^2(u):+(1-\beta^2/2)\phi''(u)+\frac{1}{2}:\xi\xi'(u):+\frac{\epsilon\beta^2}{16}\\ 
\frac{i^{1/2}\beta^2}{\sqrt{2}}G(u)&=&\phi '\xi(u)+(1-\beta^2/2)\xi '(u),
\nonumber
\end{eqnarray}
where
\begin{eqnarray}
&&\phi(u)=iQ+iPu+\sum_n\frac{a_{-n}}{n}e^{inu},\qquad
\xi(u)=i^{-1/2}\sum_n\xi_ne^{-inu},\\
&&[Q,P]=\frac{i}{2}\beta^2 ,\quad 
[a_n,a_m]=\frac{\beta^2}{2}n\delta_{n+m,0},\qquad
\{\xi_n,\xi_m\}=\beta^2\delta_{n+m,0}.\nonumber
\end{eqnarray}
Recall that there are two types of boundary conditions on 
$\xi$: $\xi(u+2\pi)=\pm\xi(u)$. The sign ``+'' corresponds 
to the R sector,the case
when $\xi$ is integer modded, the ``--'' sign corresponds to the NS sector and
$\xi$ is half-integer modded. The variable $\epsilon$ in (13)
is equal to zero
in the R case and equal to 1 in the NS case.\\
One can expand $T(u)$ and $G(u)$ by modes in such a way: $
T(u)=\sum_nL_{-n}e^{inu}-\frac{\hat{c}}{16}$, $G(u)=\sum_nG_{-n}e^{inu}
$,
where  $\hat{c}=5-2(\frac{\beta^2}{2}+\frac{2}{\beta^2})$  and $L_n,G_m$ 
generate the superconformal algebra:
\begin{eqnarray}
[L_n,L_m]&=&(n-m)L_{n+m}+\frac{\hat{c}}{8}(n^3-n)\delta_{n,-m}\\
\lbrack L_n,G_m\rbrack&=&(\frac{n}{2}-m)G_{m+n}\nonumber\\
\lbrack G_n,G_m\rbrack&=&2L_{n+m}+\delta_{n,-m}\frac{\hat{c}}{2}(n^2-1/4).
\nonumber
\end{eqnarray}
In the classical limit  $c\to -\infty$ (the same is $\beta^2\to 0$) 
the following substitution:
$
T(u)\to-\frac{\hat{c}}{4}U(u)$, 
$G(u)\to-\frac{\hat{c}}{2\sqrt{2i}}\alpha(u)$,
$[,]\to \frac{4\pi}{i\hat{c}}\{,\}
$
reduce the above algebra to the Poisson bracket algebra of 
super-KdV theory.\\ 
\hspace*{5mm}Let now $F_p$ be the highest weight 
module over the oscillator algebra of 
$a_n$, $\xi_m$ with the highest weight vector (ground state) $|p\rangle$ 
determined by the 
eigenvalue of $P$ and nilpotency condition of the action of the positive modes:
\begin{eqnarray}
P|p\rangle=p|p\rangle,\quad 
a_n|p\rangle=0, \quad \xi_m|p\rangle=0\quad n,m > 0.
\end{eqnarray}
In the case of the R sector the highest weight becomes doubly degenerate
due to the presence of zero mode $\xi_0$. So, there are two ground states
$|p,+\rangle$ and $|p,-\rangle$: $|p,+\rangle = \xi_0|p,-\rangle$.
Using the above free field representation of the superconformal algebra
one can obtain that for generic $\hat{c}$ and $p$, $F_p$ is isomorphic to the 
super-Virasoro module with the highest weight vector $|p\rangle$:
\begin{eqnarray}
L_0|p\rangle&=&\Delta_{NS}|p\rangle,\quad \Delta_{NS}=
\Bigg(\frac{p}{\beta}\Bigg)^2 + \frac{\hat{c}-1}{16}
\end{eqnarray}
in the NS sector and module with two highest weight vectors in the Ramond case:
\begin{eqnarray}
L_0|p,\pm\rangle=\Delta_{R}|p,\pm\rangle,\quad\Delta_{R}=
\Bigg(\frac{p}{\beta}\Bigg)^2 + \frac{\hat{c}}{16},\quad
|p,+\rangle=\frac{\beta^2}{\sqrt{2}p}G_0|p,-\rangle.
\end{eqnarray}
The space $F_p$, now considered as super-Virasoro module, splits in the sum of 
finite-dimensional subspaces, determined by the value of $L_0$: $
F_p=\oplus^{\infty}_{k=0}F_p^{(k)}$, $L_0 F_p^{(k)}=(\Delta + k) F_p^{(k)}$.
The quantum versions of local integrals of motion should act invariantly on 
the subspaces $F_p^{(k)}$. Thus, the diagonalization
of IM reduces (in a given subspace $ F_p^{(k)}$) to the finite purely 
algebraic problem,
which however rapidly become very complex for large $k$. It should 
be noted also that in the case of the 
Ramond sector $G_0$ does not commute with IM (even classically), 
so IM mix $|p,+\rangle$ and $|p,-\rangle$.\\
    At the end of this section we introduce another useful notion 
-- vertex operator.
We need two types of them: 
$V_B^{(a)}=\int \d\theta\theta:e^{a\Phi}:$ (''bosonic'') and 
$V_F^{(b)}=\int \d\theta:e^{b\Phi}:$ (``fermionic''),
where $\Phi(u,\theta)=\phi(u)-\theta\xi(u)$ is a superfield. Thus,
$
V_B^{(a)}=:e^{a\phi}:,\quad V_F^{(b)}=-b\xi:e^{b\phi}:
$
and normal ordering here means that
$:e^{c\phi(u)}:=
\exp\Big(c\sum_{n=1}^{\infty}\frac{a_{-n}}{n}e^{inu}\Big)
\exp\Big(ci(Q+Pu)\Big)\exp\Big(-c\sum_{n=1}^{\infty}\frac{a_{n}}{n}e^{-inu}
\Big).$
\section{Quantum Monodromy Matrix and Fusion Relations}
 In this section we will construct the quantum versions of monodromy matrices,
operators $\mathbf{L}_s$ and $\mathbf{t}_s$.\\
   The classical monodromy matrix is based on the $\widehat {osp}(1|2)$ 
affine Lie algebra.
In the quantum case the underlying algebra is quan\-tum 
$\widehat {osp}_q(1|2)$ \cite{Tolst} with $q=e^{i\pi\beta^2}$
and generators, corresponding to even root $\alpha_0$ and odd
root $\alpha$:
\begin{eqnarray} 
&&[h_\gamma,h_{\gamma'}]=0\quad(\gamma,\gamma'=\alpha, d, \alpha_0),\quad
 [e_\beta,e_{\beta'}]=\delta_{\beta,\beta'}[h_\beta]\quad(\beta,\beta'=\alpha,
\alpha_0),\nonumber\\
&&[h_d,e_{\pm\alpha_0}]=\pm e_{\pm\alpha_0},\quad
  [h_d,e_{\pm\alpha}]=0,\quad
  [h_{\alpha_0},e_{\pm\alpha_0}]=2e_{\pm\alpha_0},\\
&&[h_{\alpha_0},e_{\pm\alpha}]=\mp e_{\mp\alpha},\quad
  [h_{\alpha},e_{\pm\alpha}]=\pm\frac{1}{2} e_{\pm\alpha},\quad
  [h_{\alpha},e_{\pm\alpha_0}]=\mp e_{\alpha_0},\nonumber\\
&&[[e_{\pm\alpha},e_{\pm\alpha_0}]_q,e_{\pm\alpha_0}]_q=0,\quad
  [e_{\pm\alpha},[e_{\pm\alpha},[e_{\pm\alpha}[e_{\pm\alpha},[e_{\pm\alpha},
e_{\pm\alpha_0} ]_q]_q]_q]_q]_q=0.\nonumber
\end{eqnarray}
Here $[,]_q$ is the super q-commutator: $[e_a,e_b]=
e_ae_b-q^{(a,b)}(-1)^{p(a)p(b)}e_be_a$ and parity $p$ is defined as 
follows: $p(h_{\alpha_0})=0,\quad 
p(h_{\alpha})=0,\quad p(e_{\pm\alpha_0})=0 ,
\quad p(e_{\pm\alpha})=1$.
Also, as usual, $[h_{\beta}]=\frac{q^{h_\beta}-q^{-h_\beta}}{q-q^{-1}}$.
The finite dimensional representations $\pi_s^{(q)}(\lambda)$ of 
$\widehat {osp}_q(1|2)$ 
can be characterized by integer number s.
The triple $h_{\alpha_0}$, $e_{\pm\alpha_0}$ forms the $sl_q(2)$
subalgebra and the whole $\widehat{osp}_q(1|2)$ finite dimensional
irreducible representation is decomposed into the direct sum:
$\pi_s^{(q)}(\lambda)=\oplus_{j=0}^{[s/2]}\mu^{(q)}_{j}$, 
where $\mu^{(q)}_{j}$ are the representations of $sl_q(2)$ 
with spin $j$ ($j$ runs through integer and half integer numbers). 
The odd generators $e_{\pm\alpha}$ acting
on $\mu^{(q)}_{j}$ mix $\mu^{(q)}_{j+1/2}$ and $\mu^{(q)}_{j-1/2}$ 
representations.
In the classical limit ($q \to 1$) the representation
$\pi_s^{(q)}(\lambda)$ splits into the direct sum of the
irreducible representations $\pi_r(\lambda)$ of $\widehat{osp}(1|2)$:
$
\pi_s^{(1)}(\lambda)=\oplus_{k=0}^{[s/2]}\pi_{s-2k}(\lambda).
$
In this sum k runs through integer numbers.
The structure of irreducible finite dimensional 
representations of $\widehat {osp}_q(1|2)$ 
is similar to those of $(A_2^{(2)})_q$ \cite{2}. This is the consequence of 
the coincidence of their Cartan matrices.\\
\hspace*{5mm}
After these preparations we are ready to introduce the quantum analogue of
$\mathbf{L}_s$ operators:
\begin{eqnarray}
\mathbf{L}_s^{(q)}&=&\pi_s^{(q)}(\lambda)(\mathbf{L}^{(q)})\\
&=&\pi_s^{(q)}(\lambda)\Bigg(e^{-i\pi Ph_{\alpha_0}}P\exp\bigg(\int_0^{2\pi} \d u
\Big(:e^{2\phi(u)}:e_{\alpha_0}
+\xi(u):e^{-\phi(u)}:e_{\alpha}\Big)\bigg)\Bigg).\nonumber
\end{eqnarray}
One can see that the term $e^{-2\phi(u)}2e_{\alpha}^2$ is 
missing in the P-exponent in comparison
with the classical case (11). This is the general 
result for superalgebras and we 
will return to this in \cite{progr}.
Analyzing the singularity properties of the integrands in P-exponent
of $\mathbf{L}^{(q)}_s(\lambda)$ one can find that the integrals are 
convergent for
$
-\infty<\hat c< 0
$
and need regularization for a wider region.\\
\hspace*{5mm}
Now let's prove that in the classical limit $\mathbf{L}^{(q)}$ will 
coincide with $\mathbf{L}$.
  First let's analyse the products of the operators we have in the P-exponent.
The product of the two fermion operators can be written in such a way:
\begin{eqnarray}
\xi(u)\xi(u')=:\xi(u)\xi(u'):-i\beta^2 
\frac{e^{-\kappa \frac{i}{2}(u-u')}}
{e^{\frac{i}{2}(u-u')}-e^{-\frac{i}{2}(u-u')}}.
\end{eqnarray}
where $\kappa$ is equal to zero in the NS sector and equal to 1 in 
the R sector
Also, for vertex operators we have:
\begin{eqnarray}
:e^{a\phi(u)}::e^{b\phi(u')}:=
(e^{\frac{i}{2}(u-u')}-e^{-\frac{i}{2}(u-u')})^{\frac{ab\beta^2}{2}}
:e^{a\phi(u)+b\phi(u')}:,
\end{eqnarray}
We can rewrite these products, extracting the singular parts:
\begin{eqnarray}
&&\xi(u)\xi(u')=
-\frac{i\beta^2}{(iu-iu')}+\sum_{k=1}^{\infty}c_k(u)(iu-iu')^k,\\
&&:e^{a\phi(u)}::e^{b\phi(u')}:=(iu-iu')^{\frac{ab\beta^2}{2}}
(:e^{(a+b)\phi(u)}:+\sum_{k=1}^{\infty}d_k(u)(iu-iu')^k),
\end{eqnarray}
where $c_k(u)$ and $d_k(u)$ are operator-valued functions of $u$.
Now the  $\mathbf{L}^{(q)}(\lambda)$ operator.
can be expressed in the following way:
\begin{eqnarray}
\mathbf{L}^{(q)}&=&
e^{-i\pi Ph_{\alpha_0}}\lim_{N\to\infty}\prod_{m=1}^{N}\tau_m^{(q)},\qquad
\tau_m^{(q)}=P\exp\int_{x_{m-1}}^{x_{m}}\d u K(u),\\
K(u)&\equiv&:e^{2\phi(u)}:e_{\alpha_0}+
\xi(u):e^{-\phi(u)}:e_{\alpha}. \nonumber
\end{eqnarray}
Here we have divided  the interval $[0,2\pi]$ into small 
intervals $[x_m,x_{m+1}]$
with $x_{m+1}-x_m=\Delta=2\pi/N$. 
Studying the behaviour of the first two iterations when $\beta^2\to 0$:
\begin{eqnarray}
\tau_m^{(q)}=1+\int_{x_{m-1}}^{x_{m}}\d u K(u) +
\int_{x_{m-1}}^{x_{m}}\d u K(u)\int_{x_{m-1}}^{u}\d u'K(u')+
O(\Delta^2). 
\end{eqnarray}
we conclude that the second iteration
can give contribution to the first one. To see this let's  consider the 
expression that comes from the second iteration:
\begin{eqnarray}
-\int_{x_{m-1}}^{x_{m}}\d u\xi(u)\int_{x_{m-1}}^{u}\d u'\xi(u'):e^{-\phi(u)}:
:e^{-\phi(u')}: e_{\alpha}^2.
\end{eqnarray}
Now, using the above operator products and seeking the terms of order
$\Delta^{1+\beta^2}$ (only those can give us the first iteration terms in 
$\beta^2\to 0$ limit) one obtains that their contribution is:
\begin{eqnarray}
&&i\beta^{2}\int_{x_{m-1}}^{x_{m}}\d u\int_{x_{m-1}}^{u}\d u'
(iu-iu')^{\frac{\beta^2}{2}-1}:e^{-2\phi(u)}: e_{\alpha}^2\\
&=&2\int_{x_{m-1}}^{x_{m}}\d u:e^{-2\phi(u)}:(iu-ix_{m-1})^{\frac{\beta^2}{2}}
e_{\alpha}^2.\nonumber
\end{eqnarray}
Considering this in the classical limit we 
recognize the familiar terms from $\mathbf{L}$:
\begin{eqnarray}
\tau_m^{(1)}=1+\int_{x_{m-1}}^{x_{m}}\d u\Big(\xi(u)e^{-\phi(u)}e_{\alpha}
+ e^{2\phi(u)}e_{\alpha_0} +
e^{-2\phi(u)} 2 e_{\alpha}^2\Big)+O(\Delta^2).
\end{eqnarray}
 Collecting all
$\tau_m^{(1)}$ one obtains the desired result:
$
\mathbf{L}^{(1)}=\mathbf{L}.
$
Using information about representations one can also get: 
$
\mathbf{L}_s^{(1)}(\lambda)=\sum_{k=0}^{[s/2]}\mathbf{L}_{s-2k}(\lambda),
$
where $k$ runs over integer numbers.\\
\hspace*{5mm}Using the properties of quantum R-matrix \cite{leshouches} one 
obtains
that $\mathbf{R}\Delta(\mathbf{L}^{(q)})=\Delta^{op}(\mathbf{L}^{(q)})
\mathbf{R}$, where $\Delta$ and $\Delta^{op}$ are coproduct and 
opposite coproduct of $\widehat{osp}_q(1|2)$ \cite{Tolst} correspondingly.
Factorizing $\Delta(\mathbf{L}^{(q)})$ and 
$\Delta^{op}(\mathbf{L}^{(q)})$ ,  
according to the properties of vertex operators and P-exponent, 
we get the so called RTT-relation 
\cite{leshouches},\cite{sklyan}:
\begin{eqnarray}
&&\mathbf{R}_{ss'}(\lambda\mu^{-1})
\Big(\mathbf{L}_s^{(q)}(\lambda)\otimes \mathbf{I}\Big)\Big(\mathbf{I}
\otimes \mathbf{L}_{s'}^{(q)}(\mu)\Big)\\
&&=(\mathbf{I}\otimes \mathbf{L}_{s'}^{(q)}(\mu)\Big)
\Big(\mathbf{L}_s^{(q)}(\lambda)\otimes \mathbf{I}\Big)\mathbf{R}_{ss'}
(\lambda\mu^{-1}), \nonumber
\end{eqnarray}
where $\mathbf{R}_{ss'}$ is the trigonometric solution of the corresponding 
Yang-Baxter equation \cite{Kulish} which acts in the space $\pi_s(\lambda)
\otimes\pi_{s'}(\mu)$.\\
\hspace*{5mm}Let's define now the ``transfer matrices'' which are 
the quantum analogues of
the traces of monodromy matrices: $ 
\mathbf{t}_s^{(q)}(\lambda)=str\pi_s(\lambda) (e^{-i\pi ph_{\alpha_0}}
\mathbf{L}_s^{(q)})
$.
According to the RTT-relation one obtains:
\begin{eqnarray}\label{eq:qTinv}
[\mathbf{t}_s^{(q)}(\lambda),\mathbf{t}_{s'}^{(q)}(\mu)]=0.
\end{eqnarray}
Considering the first nontrivial representation (s=1) it is easy 
to find the expression for $\mathbf{t}_1^{(q)}(\lambda)\equiv
\mathbf{t}^{(q)}(\lambda)$: $\mathbf{t}^{(q)}(\lambda)=
1-2\cos(2\pi iP) + \sum^{\infty}_{n=1}\lambda^{2n}Q_n$, 
where $Q_n$ are nonlocal conservation laws, which (with the use
of (\ref{eq:qTinv})) are mutually commuting:
$[Q_n,Q_m]=0$.
Following [1],[2] we expect also that $\mathbf{t}^{(q)}(\lambda)$ 
generates local IM as in the 
classical case. Using (\ref{eq:qTinv}) again one obtains, expanding
log($\mathbf{t}^{(q)}(\lambda)$):
$ 
[Q_n,I^{(q)}_{2k-1}]=0,$ $[I^{(q)}_{2l-1},I^{(q)}_{2k-1}]=0.
$
The first few orders of expansion in $\lambda^2$ of 
$\mathbf{t}_s^{(q)}(\lambda)$ results in the following fusion
relation :
\begin{eqnarray}\label{eq:fusion}
\mathbf{t}^{(q)}_s(q^{1/4}\lambda)\mathbf{t}^{(q)}_s(q^{-1/4}\lambda)=
\mathbf{t}^{(q)}_{s+1}(q^{\frac{1}{2\beta^2}}\lambda)
\mathbf{t}^{(q)}_{s-1}(q^{\frac{1}{2\beta^2}}\lambda)+
\mathbf{t}^{(q)}_s(\lambda).
\end{eqnarray}
This result also reminds the fusion relation for 
$(A_2^{(2)})_q$ case \cite{2}. 
\section{Discussion}
Returning to the quantum Miura transformation (13) 
it should be noted that one can choose another version:
\begin{eqnarray}
&&-\beta^2T(u)=:\phi'^2(u):-(1-\beta^2)\phi''(u)+\frac{1}{2}:\xi\xi'(u):+
\frac{\epsilon\beta^2}{16}.
\end{eqnarray}
The reason why we introduce this one is the following:
we have two candidates to perturb the model without spoiling the conservation 
laws, the so-called ``integrable perturbations'' \cite{int}:
\begin{eqnarray} \label{eq:superToda}
V_1=\int \d\theta\int^{2\pi}_{0}\d u e^{-\Phi}, \quad
V_2=\int \d\theta\int^{2\pi}_{0}\d u \theta e^{2\Phi}.
\end{eqnarray} 
Operator $V_1$ is screening for the 
deformation (13), the dimension of other one is
$h_{1,5}(\hat c)-1/2$. But one should be able also to consider $V_2$ as 
screening -- that is one should use (33). In this case the dimension of 
other one is $1/2+h_{1,2}(c)$ where $c=13-6(\beta^2 + 1/\beta^2)$ is the
central charge of Virasoro algebra, generated by (33) without fermion
term.
 When one of these operators is screening, another one is chosen as
a perturbation, so one can relate the obtained model with 
$\widehat {osp}(1|2)$ supersymmetric Toda field theory \cite{Olsh} with
the corresponding Lagrangian:
$
\mathcal{L}=
D_{u,\theta}\Phi D_{\bar{u},\bar{\theta}}\Phi - e^{-\Phi} -
\theta\bar{\theta} e^{2\Phi}.
$\\ 
\hspace*{5mm}
Really, when $\beta^2$ is rational the obtained model can be treated as
superconformal minimal model, 
perturbed by $h_{1,5}(\hat c)-1/2$ dimensional operator,
or as minimal model, perturbed by $1/2+h_{1,2}(c)$ dimensional operator.
We conjecture, that the same one could obtain from the quantum group 
reduction of the theory
with the mentioned Lagrangian.\\
\hspace*{5mm}The solution of the functional system of equations (32) due 
to the conjecture of [1] determines the whole set of eigenvalues of 
$\mathbf{t}_s^{(q)}(\lambda)$ in the model. In the case when q is root of 
unity, this system should become a closed system of equations 
(quantum group truncation). Also, due to the results of \cite{1} and \cite{2} 
one can suppose that these functional equations can be transformed 
to the so-called Thermodynamic Bethe Ansatz equations \cite{TBA}, giving 
the description of the ``massless S-matrix'' theory associated with minimal 
superconformal field theories.

\section*{Acknowledgments}
We are grateful to P.I. Etingof, M.A. Semenov-Tian-Shansky, 
F.A. Smirnov and V.O. Tarasov for useful discussions.

\end{document}